\documentclass[aps,prd,twocolumn,amsfonts,amssymb,amsmath,showpacs,letterpaper,superscriptaddress,nofootinbib,orcidlink]{revtex4-2} 

\usepackage{graphicx} 
\usepackage{dcolumn}
\usepackage{bm} 
\usepackage{mathtools}
\usepackage[utf8]{inputenc}
\usepackage[T1]{fontenc}
\usepackage{booktabs, array, float, tabularx, lipsum, multirow}
\usepackage{amssymb} 
\usepackage{siunitx, xcolor}
\usepackage[version=4]{mhchem}
\graphicspath{{figs/}{figsgaoerb/}} 
\usepackage[citecolor=blue,linkcolor=blue,colorlinks=true,urlcolor=blue]{hyperref}
\usepackage{hyperxmp} 

\newcommand{\orcid}[1]{%
  \href{https://orcid.org/#1}{\includegraphics[height=1.8ex]{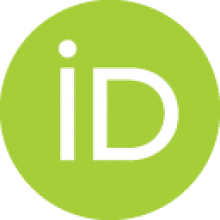}}%
}

\begin{document}
\preprint{APS/123-QED}
\title{\textbf{Lense-Thirring Precession Modulates Repeated Lensing of Continues Gravitational Wave Sources in AGN Disks } 
}%

\author{Yu-Zhe Li\,\orcid{0009-0002-9215-5618}}
\email{U202215867@hust.edu.cn}
\affiliation{Department of Astronomy, School of Physics, Huazhong University of Science and Technology, Luoyu Road 1037, Wuhan, 430074, China}

\author{Wen-Long Xu\,\orcid{0009-0003-9792-9325}}
\affiliation{Department of Astronomy, School of Physics, Huazhong University of Science and Technology, Luoyu Road 1037, Wuhan, 430074, China}

\author{Yi-Gu Chen\,\orcid{0000-0002-8043-6650}}
\affiliation{Department of Astronomy, School of Physics, Huazhong University of Science and Technology, Luoyu Road 1037, Wuhan, 430074, China}

\author{Wei-Hua Lei\,\orcid{0000-0003-3440-1526}}
\email{leiwh@hust.edu.cn}
\affiliation{Department of Astronomy, School of Physics, Huazhong University of Science and Technology, Luoyu Road 1037, Wuhan, 430074, China}

\date{\today}

\begin{abstract}
Gravitational lensing of gravitational waves (GWs) offers a novel observational channel that complements traditional electromagnetic approaches and provides unique insights into the astrophysical environments of GW sources. In this work, we investigate the repeated lensing of continuous gravitational wave (CW) sources in active galactic nucleus (AGN) disks by central supermassive black holes (SMBHs), focusing on the imprint of SMBH spin via the Lense-Thirring (LT) effect. Although typically weak and challenging to observe, the spin-induced precession of source orbits can accumulate over time, thereby modulating the lensing geometry. Such modulations influence the magnification, duration, and waveform structure of each repeated lensing event, and enhance the overall probability of lensing occurrences. Using matched filtering, we demonstrate that spin-dependent signatures may be detectable, suggesting that if such repeated lensing of CW signals were to be observed in the future, the framework developed here would offer a means to interpret the associated modulation features and to constrain the underlying system parameters.

\end{abstract}

\date[\relax]{Dated: \today }

\maketitle


\section{\label{sec:level1}INTRODUCTION }
Gravitational lensing has emerged as a powerful and indispensable tool in astrophysics and cosmology. It plays a central role in detecting dark matter, constraining the nature and environments of compact objects, and mapping the large-scale structure of the Universe \cite{mandelbaumWeakLensingPrecision2018,liaoStronglyLensedTransient2022,bartelmannWeakGravitationalLensing2001,bartelmannGravitationalLensing2010}.

Previous studies of gravitational lensing focused primarily on electromagnetic (EM) messengers. However, since LIGO's first detection of gravitational waves (GW) from a binary black hole merger (GW150914) in 2015 \citep{2016PhRvL.116m1103A,2016PhRvL.116f1102A}, GW astronomy and the lensing of GWs, particularly those from compact binary coalescence, have emerged as a vibrant research frontier. The unique low-frequency properties, coherence nature, and characteristic detection methods of GWs lead to fundamentally different lensing phenomena compared to those of EM waves. These distinctive features make wave-optical effects (especially diffraction) significant for low-frequency GWs and massive black holes \cite{Takahashi_2003}. In EM observations, lensing manifests primarily as image separations in space, while in GW signals, it appears as waveform distortions in time, making phase effects particularly significant \cite{ezquiagaPhaseEffectsStrong2021}. This difference has sparked ongoing discussions about how to classify GW lensing into strong and microlensing regimes \cite{paczynskiGravitationalMicrolensingGalactic1986,aliDetectabilityStronglyLensed2023,suyamprakasamMicrolensingLongdurationGravitational2025}. While no definitive detection of lensed GWs has been reported to date \cite{abbottSearch2024,Abbott_Search2021,hannukselaSearch2019}, the advent of next-generation ground- and space-based detectors promises to greatly enhance the prospects for such observations \cite{ngPrecise2018,sesanaProspects2016}, potentially enabling the use of GW lensing as a novel probe of the environments surrounding their sources \cite{ubachConstrainingEnvironmentCompact2025a,grespanStrongGravitationalLensing2023}.

In recent years, GW sources embedded in AGN environments, as well as the lensing of GWs therein, have emerged as new focal points and promising contributors in GW research \cite{2025ApJ...986...61L}. This growing interest is driven by their potential to enhance the formation rate of compact objects as GW sources \cite{fordBinary2022,secundaOrbital2019,rowanBlack2023}, and by the fact that the central supermassive black hole (SMBH) can act as a strong gravitational lens for sources located behind it. A vital component of AGN systems is the SMBH, whose gravitational influence on nearby GW sources has been the focus of numerous previous studies. These include the impact of the SMBH on the orbital dynamics of GW sources through mechanisms such as Lidov–Kozai (LK) oscillations and de Sitter (dS) precession \cite{yuDetectingGravitationalLensing2021,yangHierarchical2019}, as well as its role as a gravitational lens, with corresponding estimates of lensing probabilities, event rates \cite{leong_constraining_2025,xuEstimating2025}, and the repeated lensing of continuous GW (CW) signals \cite{dorazio_repeated_2020}.

However, the effects of SMBH spin have received comparatively poor attention. Most existing studies neglect spin, primarily due to its relatively weak and short-range influence, as well as the significant challenges associated with its measurement \cite{reynoldsObservational2021}. The study \cite{fangImpactSpinningSupermassive2019} investigates the impact of frame-dragging effects induced by the spin of SMBH on GW sources and how these effects manifest in the waveforms. However, detecting such signatures requires prolonged continuous observations and high signal-to-noise ratios (SNR). Although the spin effects are subtle, gravitational lensing is highly sensitive to the position of the source, especially when the source, lens, and observer are well aligned. This motivates the consideration of using lensing phenomena to reveal the weak spin-induced Lense-Thirring (LT) effect.
Since the LT effect is relatively weak, it requires time to accumulate. Therefore, we focus on CW sources in AGN disks rather than transient compact binary coalescence (CBC). We mainly consider stellar-mass compact binaries (BBHs) within the LISA band as potential CW sources, which are expected to be detectable by future space-based detectors \cite{wagg_gravitational_2022, amaroseoane2017lisa}. We also discuss rapidly rotating neutron stars \cite{sieniawskaContinuousGravitationalWaves2019}, and boson clouds around spinning stellar Black Holes as CW sources for ground-based detectors \cite{rilesSearches2023}; however, neutron stars beyond our Galaxy are unlikely to be detectable current or foreseeable ground-based detectors. Moreover, since the LT precession period is much longer than the orbital period, repeated lensing becomes necessary; this is enabled by the CW source orbiting the SMBH and repeatedly crossing the Einstein radius. The orbital angular momentum vector of the CW source undergoes precession around the spin axis of SMBH. As a result, an edge-on orbit of the CW source may become misaligned, and the orbit might entirely avoid entering the Einstein radius. Conversely, orbits initially outside the lensing region can be dragged into it due to this precession. We investigate the lensing duration, magnification, and apply matched filtering to the waveforms to detect the imprint of spin.

This paper is organized as follows. In Sec.~\ref{sec:GLofCW}, we introduce the basic theory of GW lensing, describe the characteristics of CW sources, and discuss how such sources are affected by lensing. Sec.~\ref{sec:LTandspace} presents the manifestation of the LT effect in our system and examines the contributions from other dynamical effects. In Sec.~\ref{sec:LTonlense}, We investigate how LT-induced precession modifies the properties of lensed CW signals, analyze waveform characteristics and their detectability, explore the dependence on GW frequency, spin magnitude, and geometric configuration, and further study the potential enhancement in lensing probability. Conclusions and discussions are given in Sec.~\ref{sec:discussion}.

\section{\label{sec:GLofCW}Gravitational lensing of CW source in AGN disc}

In this section, we discuss the gravitational lensing features of CWs. We begin by briefly reviewing the fundamentals of gravitational lensing and its key characteristics (Sec.~\ref{subsec:GL}). Next, we focus on the specific lensing signatures associated with CWs and examine how variations in the source position of CWs can affect the resulting lensing signals (Sec.~\ref{subsec:CW}).
\begin{figure}[htbp]
  \centering
  \includegraphics[width=0.45\textwidth]{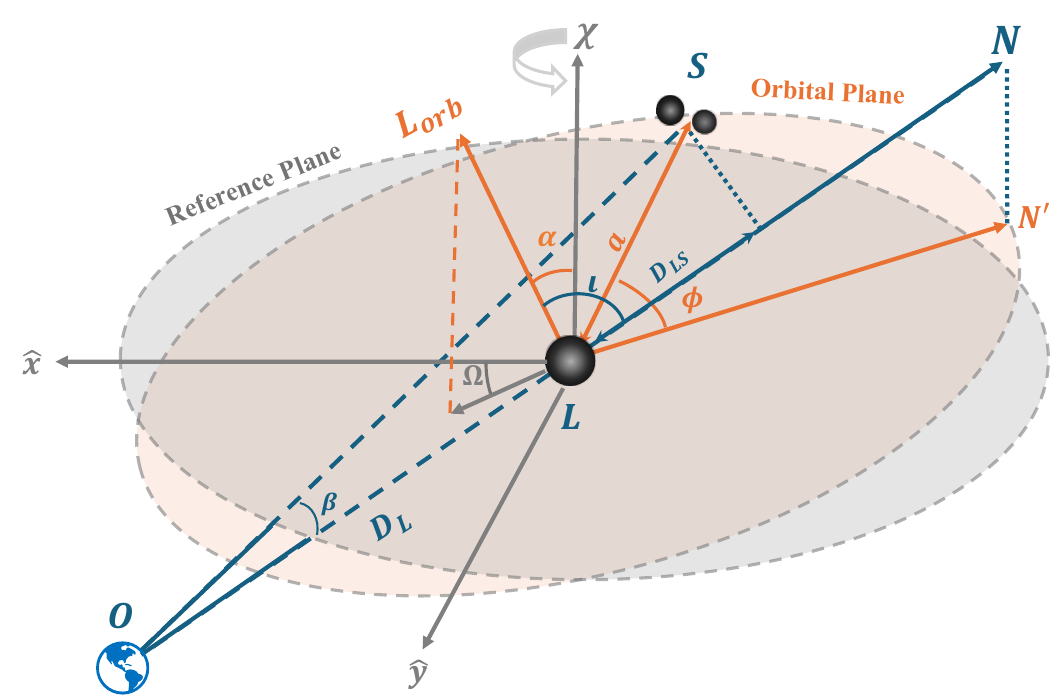}
  \caption{Geometry of the SMBH–CW source system. We define the $z$-axis to be aligned with the spin vector $\boldsymbol{\chi}$ of the SMBH. The reference plane is then the plane perpendicular to this spin direction. The CW source resides in the orbital plane with an orbital radius $a$. The orbital angular momentum $\boldsymbol{L}_{\mathrm{orb}}$ forms an angle $\alpha$ with $\boldsymbol{\chi}$. The observer’s line of sight is denoted by $\boldsymbol{N}$, and its projection onto the reference plane is $\boldsymbol{N}'$. The source direction is $\boldsymbol{S}$, located on the AGN disk. The angle between $\boldsymbol{N}$ and $\boldsymbol{L}_{\mathrm{orb}}$ is defined as $\iota$; the angle between $\boldsymbol{S}$ and $\boldsymbol{N}'$ is $\phi$; and the source position is $\beta$. A Cartesian coordinate system is defined in the reference plane with $z$-axis along $\boldsymbol{\chi}$; note that this differs from the \emph{orbital-plane} coordinate system where $z$ is along $\boldsymbol{L}_{\mathrm{orb}}$. The projection of $\boldsymbol{L}_{\mathrm{orb}}$ onto the reference plane defines an angle $\Omega$ with respect to the $x$-axis. $D_{\rm L}$ and $D_{\rm LS}$ denote the distances from the lens to the observer and the projected distance from the lens to the source along $\boldsymbol{N}$, respectively.}
  \label{fig:example}
\end{figure}
\subsection{\label{subsec:GL}Gravitational lensing}
We first introduce an \emph{orbital-plane} coordinate system, which is distinct from the spin reference frame shown in Fig.~\ref{fig:example}. In this frame, the $Z$-axis is aligned with the orbital angular momentum of the CW source, while the $X$--$Y$ plane coincides with the source's orbital plane (see Fig.1 in Ref. \cite{leong_constraining_2025}). This choice simplifies the description of the source's position. In contrast, the coordinate system in Fig.~\ref{fig:example} adopts the $Z$-axis along the spin angular momentum of the SMBH, which is more convenient for illustrating the precession of the orbit relative to the SMBH spin axis.
In the \emph{orbital-plane} coordinate frame, the unit vector pointing toward the source is given by
\begin{equation}
   \hat{S} = \left( \cos \phi,\ \sin \phi,\ 0 \right), 
\end{equation}
where $\phi$ source orbital phase $\phi=\Omega_{\rm o}t$ and $\Omega_{\rm o} = \sqrt{\frac{GM_{\rm L}}{a^3}}$ is the angular orbital frequency of the outer orbit. Here, $M_{\rm L}$ denotes the mass of the SMBH, and $a$ is the outer orbital radius. In this work, we assume that CW sources reside on nearly circular orbits. This assumption is motivated by the interaction between the sources and the AGN disk: as objects evolve within or repeatedly transit through the disk, dynamical gas drag and momentum exchange tend to decrease the orbital inclination with respect to the disk plane and reduce their orbital eccentricities \cite{bartosRapid2017,fabjAligning2020,macleodEffect2020,rowanHydrodynamic2025}, driving the orbits toward a quasi-circular configuration. Although a subset of recently captured sources may occupy more eccentric or otherwise atypical orbits, we do not consider them as the primary focus of this study.

The optical axis, defined as the direction from the observer to the source, is given by
\begin{equation}
\hat{O} = -\hat{N} = \left( -\sin \iota,\ 0,\ -\cos \iota \right),
\end{equation}
where $\iota$ is the inclination angle between the observer's line of sight and the orbital angular momentum axis.

The angle $\theta$ between the source direction and the optical axis is given by the dot product:

\begin{equation}
\cos \theta = \langle \hat{S}, \hat{O} \rangle = -\cos \phi \ \sin \iota.
\end{equation}
From the geometry illustrated in Fig.\ref{fig:example}, we obtain the relation:
\begin{equation}
D_{\rm S} \tan \beta = a \sin \theta,
\end{equation}
where $a$ is the orbital radius of CW source, $\beta$ is the real source angular position.

Since $\beta$ is a small angle ($\beta \ll 1$), we take $\tan \beta \approx \beta$.
This yields a direct relation between $\beta$ and the parameters $\phi$ and $\iota$ at a given orbital radius:
\begin{equation}
\beta \approx \frac{a\sin \theta}{D_{\rm S}}=\frac{a\sqrt{1-\cos^2\phi\sin^2\iota} }{D_{\rm S}}.
\end{equation}
This result agrees with the expression given in \cite{ChenyanbeiPhysRevLett.126.021101,leong_constraining_2025}.

To simplify the analysis, we define the source position in units of the Einstein angular radius
\begin{eqnarray}
\eta = \frac{\beta}{\theta_{\rm Ein}},
\label{eq:eta}
\end{eqnarray}
where $\theta_{\rm Ein}$ is the angular Einstein angular radius defined as
\begin{eqnarray}
\theta_{\rm Ein} = \sqrt{2R_{\rm S} \frac{D_{\rm LS}}{D_{\rm L} D_{\rm S}}},
\label{eq:einstein_angle}
\end{eqnarray}
with $D_{\rm LS}$ being the distance between the lens and the source,
and $R_{\rm S} = 2GM_{\rm L}/c^2$ the Schwarzschild radius of the lens with mass $M_{\rm L}$.

In the following analysis, we adopt the point-mass lens model (PML) for simplicity. Although, in principle, a spinning Kerr black hole should produce lensing signatures that differ from those of a non-spinning lens, previous studies have shown that the effect of spin can be treated as a small correction to the point-mass model. This correction has only a minor impact on our subsequent results. A more detailed discussion is provided in Sec.~\ref{sec:discussion}.

The effect of lensing can be described by transmission factor:
\begin{equation}
    h_{\rm L}(f) = F(f,\eta)\, h_{\rm UL}(f),
    \label{eq:TFdef}
\end{equation}
where $h_{\rm UL}(f)$ is unlensed waveform template and $h_{\rm L}(f)$ is lensed waveform.
By solving the Fresnel-Kirchhoff diffraction integral for the effective lensing potential $\psi(x) = \ln(x)$, we obtain the transmission factor for the PML including wave optics effect \cite{NakamuraPhysRevLett.80.1138,Takahashi_2003,ChenyanbeiPhysRevLett.126.021101}:
\begin{align}
    F(w,\eta) =& \exp\left\{\frac{\pi w}{4} + i \frac{w}{2}
    \left[\ln\left(\frac{w}{2}\right)-2\phi_{\rm m}(\eta)\right]\right\}\nonumber \\
    &\times \Gamma(1-i\frac{w}{2})_1F_1(i\frac{w}{2}, 1; i\frac{w\eta^2}{2}),
    \label{eq:transmission_factor}
\end{align}
where $\Gamma$ is the complex gamma function, ${_1F_1}$ is the confluent hypergeometric function of the first kind, and
\begin{equation}
w(f, M_{\rm L}) = \frac{8\pi G M_{\rm L} f}{c^3}
\label{eq:w}
\end{equation}
is the dimensionless frequency. The phase correction $\phi_{\rm m}$ is defined as
\begin{equation}
\phi_{\rm m} =
\frac{1}{2}(\eta_{\rm m} - \eta)^2 - \ln(\eta),
\label{eq:phi_m}
\end{equation}
with
\begin{eqnarray}
\eta_{\rm m} = \frac{1}{2} \left(\eta + \sqrt{\eta^2 + 4} \right).
\label{eq:eta_m}
\end{eqnarray}
In the geometrical optics limit ($w \gg 1$), the transmission factor reduces to
\begin{align}
F(f, M_{\rm L}) =& |\mu_{+}|^{1/2} - i |\mu_{-}|^{1/2} e^{2\pi i f \Delta t},
\label{eq:F_GO}
\end{align}
where $\mu_\pm$ are the image magnifications and $\Delta t$ is the time delay between the two images, given by
\begin{equation}
\mu_{\pm} = \frac{1}{2} \pm \frac{\eta^2 + 2}{2 \eta \sqrt{\eta^2 + 4}},
\label{eq:magnification}
\end{equation}
and
\begin{align}
\Delta t =
&\frac{4GM_{\rm L}}{c^3} \left[ \frac{\eta \sqrt{\eta^2 + 4}}{2} + \ln{\left( \frac{\sqrt{\eta^2 + 4} + \eta}{\sqrt{\eta^2 + 4} - \eta} \right)} \right].
\label{eq:time_delay}
\end{align}

\begin{figure}[htbp]
  \centering
  \includegraphics[width=0.45\textwidth]{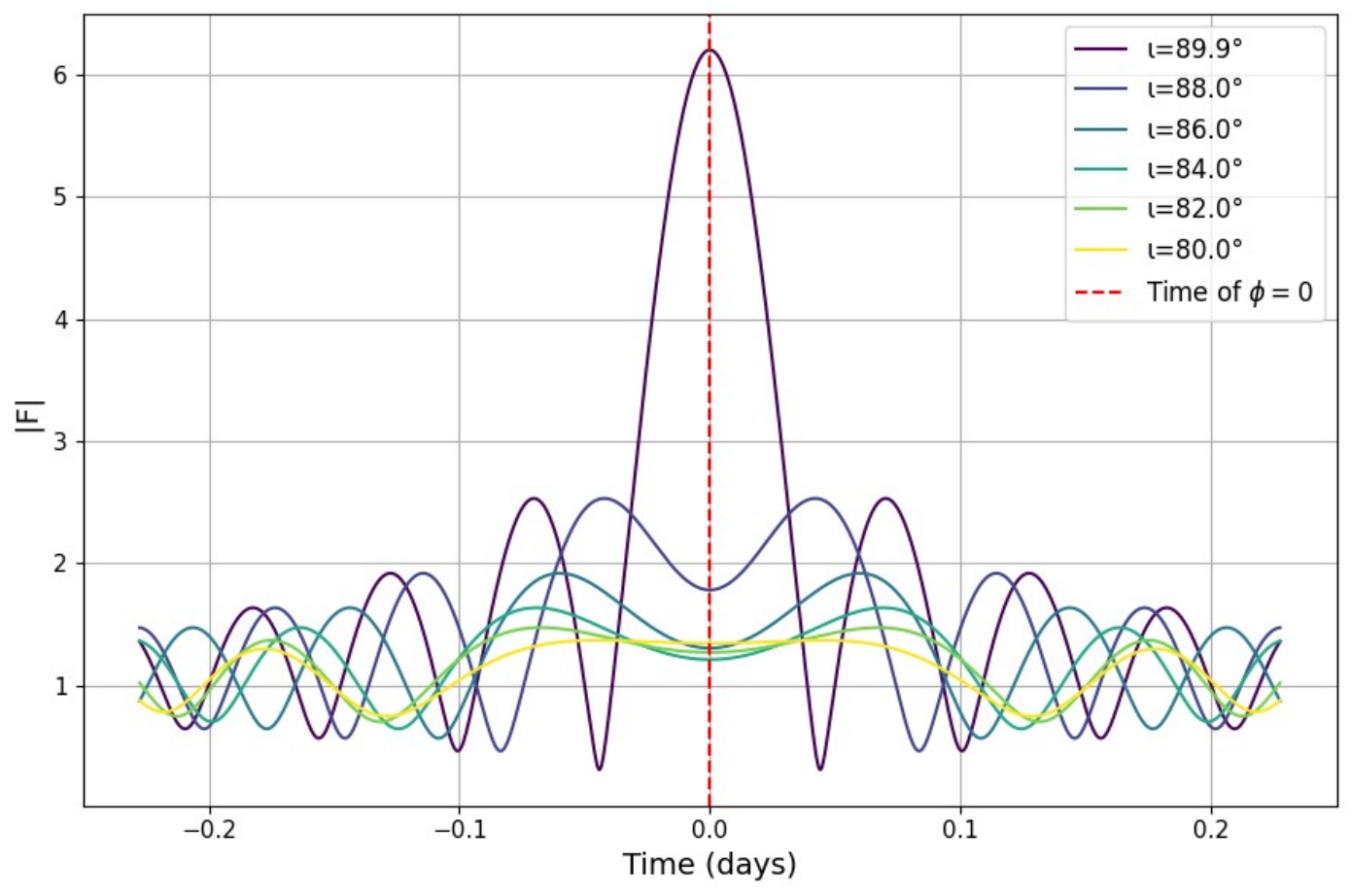}
  \caption{Variation of the transmission factor as a function of time for different inclination angles $\iota = 80^\circ$ to $89.9^\circ$ (six values), with frequency $f = 10^{-2}$ Hz ($\omega \sim 12$), where the transmission factor is computed using the full wave optics formalism Eq.\eqref{eq:transmission_factor}. The source orbits a lens of mass $M_{\rm L} = 10^7\,M_\odot$ at an orbital radius $a = 100R_{\rm s}$. The red dashed line marks the time of $\phi = 0$. The lensing effect weakens as $\iota$ deviates from $90^\circ$ due to reduced alignment.}
  \label{fig:TF001}
\end{figure}

\begin{figure}[htbp]
  \centering
  \includegraphics[width=0.45\textwidth]{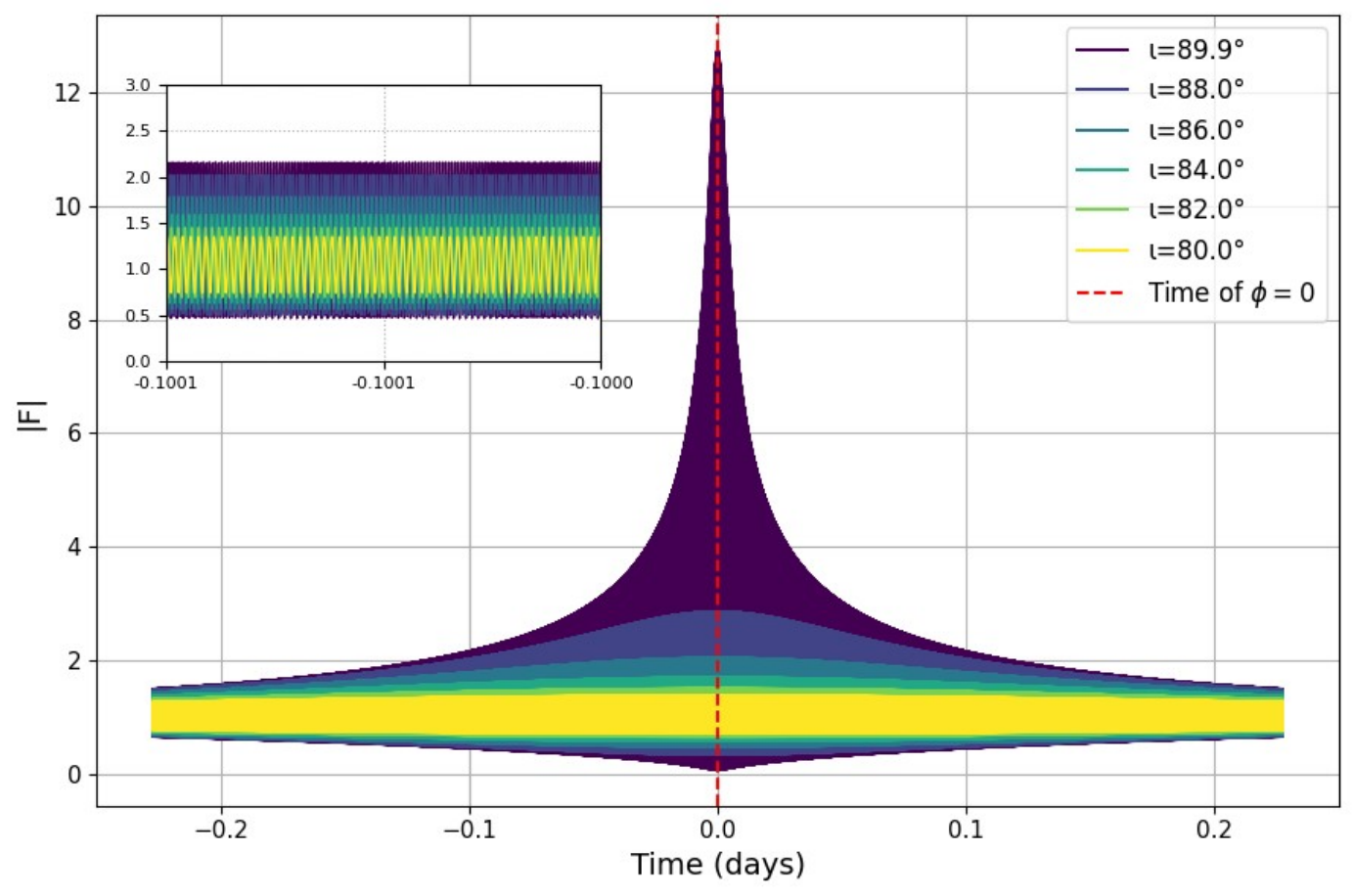}
  \caption{Same setup as Fig.~\ref{fig:TF001} but with a higher frequency of $f = 800$ Hz ($\omega \sim 10^6$), approaching the geometric optics limit. The transmission factor becomes significantly sharper and higher, exhibiting rapid oscillations, as shown in the magnified view in the upper left corner. We zoom into a narrow window of $ 1 \times 10^{-4}$ days around $-0.1$ days to clearly illustrate the fine-scale oscillatory behavior.
}
  \label{fig:TF800}
\end{figure}

In Figs. \ref{fig:TF001} and \ref{fig:TF800}, we illustrate the variation of the transmission factor caused by the orbital motion of the source as it crosses the Einstein radius. We adopt $M_{\rm L} = 10^7\,M_\odot$ and an orbital radius of $a = 100R_{\rm s}$. The time duration shown corresponds to $T_{\rm Ein}|_{\iota = 90^\circ}$, defined as the crossing time at an inclination angle $\iota = 90^\circ$. We consider six $\iota$ values from $80^\circ$ to $89.9^\circ$, and examine two representative frequencies: $10^{-2}$ Hz (corresponding to $w \sim 12$) and 800 Hz (corresponding to $w \sim 10^6$). It can be seen that as the inclination angle $\iota$ deviates from $90^\circ$, the amplitude of the lensing effect decreases. This is because $\iota = 90^\circ$ corresponds to the edge-on configuration, where a well-aligned geometry can occur at $\phi = 0$, leading to strong lensing. As $\iota$ deviates from $90^\circ$, the source crosses the Einstein radius along a chord rather than through the center, reducing the lensing effect. For sufficiently small deviations, the trajectory may even no longer intersect the Einstein radius, further diminishing the lensing signal. By comparing Figs. \ref{fig:TF001} and \ref{fig:TF800}, we observe that as the frequency increases, the transmission factor enters the geometrical optics limit, becoming increasingly sharp, and the lensing effect correspondingly more pronounced. In the well-aligned configuration, the magnification predicted by the geometrical optics approximation formally diverges (\ref{eq:magnification}).

\subsection{\label{subsec:CW}CW sources and CW lensing}
In addition to the transient GW signals generated by binary coalescences, detectors are also capable of observing long-lived signals characterized by slow frequency evolution. The sources of such signals can be broadly classified into two categories:
(1) stellar-mass compact binaries (BBH) in the early inspiral phase, which are expected to be detectable at extragalactic distances by upcoming space-based observatories such as LISA, TianQin and Taiji \cite{wagg_gravitational_2022, amaroseoane2017lisa,luoTianQin2016,huTaiji2017}; and
(2) rapidly rotating, non-axisymmetric neutron stars (NS) with ellipticity, as well as boson clouds around spinning stellar black holes are promising candidates for CW sources in the frequency band of ground-based interferometers, and are currently being searched for in LIGO data \cite{abbottGWTC3CompactBinary2023,abbottGWTC21DeepExtended2024,PhysRevD.106.102008,sieniawskaContinuousGravitationalWaves2019,abbottAllsky2022}. 
However, even within the Milky Way, population-synthesis studies predict only a very limited number of detectable neutron stars; detecting such sources in extragalactic environments would be even more challenging with current or next generation ground-based detectors \cite{cieslarDetectability2021,rilesSearches2023,pagliaroContinuousGravitationalWaves2023}. By contrast, boson clouds may in principle allow for extragalactic detectability under favorable conditions \cite{isiDirected2019,collaborationDirected2025}, but such prospects rely on highly uncertain assumptions about the boson mass, cloud formation efficiency, and source population, and therefore remain strongly speculative. It is worth noting that, in the LIGO frequency band, continuous-wave searches have already placed stringent constraints on boson condensate signals within the Milky Way, particularly for scalar bosons \cite{palombaDirect2019,abbottAllsky2022}. Moreover, the theoretically allowed mass range for bosons spans many orders of magnitude, whereas ground- and space-based gravitational-wave detectors are sensitive to only a narrow portion of this parameter space \cite{ferreiraUltralight2021,britoGravitational2017}. From this perspective, current and future searches necessarily probe a limited region of the broader landscape, and conclusions regarding detectability should therefore be interpreted with appropriate caution. A more quantitative assessment, incorporating realistic population models, cloud lifetimes, and detector sensitivities, remains an important direction for future work.

For a signal with intrinsic amplitude $\mathcal{A}$, which depends on the physical properties of the source, the signal-to-noise ratio (SNR) can be estimated as \cite{savastanoLensSgrIdentifying2024,basakProspectsObservationContinuous2023}
\begin{equation} \rho \simeq\sqrt{\frac{\langle h_{\mathrm {obs}}^{2}\rangle T_{\mathrm{obs}}}{S_{n}(f_{0})}}\simeq R \sqrt{ \frac{ \mathcal{A}^{2}}{r_{\rm s}^{2}} \frac{T_{\rm obs}}{S_{\rm n} \left( f_{0} \right)}}, \label{eq:SNR} \end{equation}
where $r_{\rm s}$ is the distance to the source, $T_{\mathrm{obs}}$ is the observation time, and $S_{\rm n}(f_{0})$ is the one-sided noise power spectral density of the detector evaluated at the signal frequency $f_{0}$.

The quantity $h_{\mathrm{obs}}$ denotes the gravitational-wave strain measured at the detector and depends on the detector response to the two polarization components of the signal. It can be written as
\begin{equation}
h_{\mathrm{obs}}(t) = F_{+}(t)\, h_{+}(t) + F_{\times}(t)\, h_{\times}(t),
\end{equation}
where $h_{+}(t)$ and $h_{\times}(t)$ are the plus and cross polarizations, respectively, and $F_{+}(t)$ and $F_{\times}(t)$ are the time-dependent antenna pattern functions, which are determined by the relative orientation between the detector and the source, as well as by the source sky location and polarization angle.

The dimensionless factor $R$ in Eq.~(\ref{eq:SNR}) represents an \emph{effective detector response}, which encapsulates the combined effects of the antenna pattern functions, the source sky position, polarization, and inclination. In practice, the detector response factor $R$ can either be taken as the
root-mean-square (RMS) average over the sky position and polarization,
or as a conservative suppression factor to account for unfavorable
source orientations. For a ground-based right-angle Michelson interferometer such as LIGO,
the standard sky- and polarization-averaged antenna response satisfies
$\langle F_+^2 \rangle = \langle F_\times^2 \rangle = 1/5$, which yields
$R_{\rm LIGO} =  \sqrt{\frac{1}{5}} \simeq 0.45 $ \cite{flanaganMeasuring1998,Moore_2015}.
This value represents an optimistic RMS response for an isotropic
distribution of sources. For LISA, whose actual arm opening angle is $60^\circ$ rather than
$90^\circ$, the response to each polarization is further rescaled by
an additional geometric factor $\sqrt{3}/2$ \cite{bertiEstimating2005}. Consequently, the
corresponding sky-averaged RMS response becomes
$R_{\rm LISA} = \frac{\sqrt{3}}{2}\sqrt{\frac{1}{5}} \simeq 0.39 .$

In realistic detectability estimates, it is often more appropriate to adopt
a conservative effective response factor, $R \simeq 0.1$--$0.3$, which
accounts for unfavorable sky locations, polarization configurations,
and time-dependent detector orientation.

It is worth emphasizing that the detector response of LISA is intrinsically more uniform over the sky than that of a single ground-based Michelson interferometer such as LIGO \cite{giampieriAntenna1997}. In the triangular configuration of LISA, the three spacecraft form an approximately equilateral triangle, and independent interferometric measurements can be constructed along each pair of arms. As a result, when integrating over time or combining multiple TDI channels, the overall LISA response to gravitational waves becomes effectively non-vanishing for almost any sky location \cite{rieggerDirectional2024}.

This is in contrast to a single LIGO-type detector, which consists of one Michelson interferometer with orthogonal arms and therefore possesses sky locations and polarization states for which the detector response vanishes \cite{liuThreedimensional2020}. Even a network of two nearly coaligned LIGO detectors (LHO and LLO) does not fully remove these blind directions, since their arm orientations were deliberately chosen to be as parallel as permitted by the Earth’s curvature.


Detecting CWs are challenging because their signals are much weaker than those from compact binary coalescences. To detect them, we need long observation times Eq. (\ref{eq:SNR}) and significant computing power for all-sky searches \cite{suyamprakasamMicrolensingLongdurationGravitational2025}. Therefore, lensing, by amplifying the signal, might play an important role in making such weak sources observable.

Over observation timescales of a few years, the frequency evolution of continuous gravitational waves (CWs) from BBHs is typically modest; for lensing timescales shorter than a day, the frequency evolution can be safely neglected. For example, a stellar-mass BBH system with component masses of $20M_\odot$ emitting at a gravitational wave frequency of $f = 10\mathrm{mHz}$—within the LISA sensitivity band—undergoes a frequency drift of about $\Delta f=0.1\mathrm{mHz}$ per year and $\Delta f=10^{-4}\mathrm{mHz}$ per day under the leading-order (0 PN) approximation, assuming gravitational wave emission as the sole energy loss mechanism \cite{PhysRev.136.B1224}.  Isolated neutron stars can be treated as quasi-monochromatic sources over observational timescales \cite{Lasky_2015,Kaplan_2009}, as their spin-down timescales—typically $\sim 5 \times 10^4$ years for gravitational wave emission and $\sim 10^8$ years for electromagnetic processes \cite{pagliaroContinuousGravitationalWaves2023,sieniawskaContinuousGravitationalWaves2019}—are much longer than the duration of current observations. Following the treatment in \cite{2019ApJ...875..139L,suyamprakasamMicrolensingLongdurationGravitational2025}, we neglect the frequency evolution of the CW signals and model them as monochromatic waves in lensing calculation.

The unlensed continuous gravitational wave (CW) signal in the frequency domain is expressed as
\begin{equation}
    {h}_{\rm UL}(f) = \mathcal{A} \cdot \delta(f - f').
\end{equation}
where $f'$ is the frequency of this CW.

When the source is in motion, the lensing parameter $\eta$ becomes time-dependent. In this case, Eq. \eqref{eq:TFdef} takes the form
\begin{equation}
F(f, \eta (t)) = \frac{{h}_{\rm L}(f,t)}{{h}_{\rm UL}(f)},
\end{equation}
and the lensed waveform is
\begin{equation}
\label{eq:lensedCW}
h_{\rm L}(t) = F(f', \eta (t))\, h_{\rm UL}(t).
\end{equation}


 Eq.~\eqref{eq:lensedCW} suggests that, under an idealized scenario where the unlensed waveform $h_{\rm UL}(t)$ is assumed to be accurately modeled or independently constrained from complementary observations, the time-dependent transmission factor $F(f', t)$ could, in principle, be inferred from the observed lensed waveform $h_{\rm L}(t)$ \cite{suyamprakasamMicrolensingLongdurationGravitational2025}. Since the instantaneous frequency $f'$ of the continuous-wave signal can generally be measured with high precision, this offers a potential avenue to probe the temporal variation of the source position $\eta(t)$ through the evolution of $F(f', t)$. In this work, we adopt this proof-of-concept approach to isolate lensing-induced modulations in a controlled setting. However, realistic searches for previously unknown sources require scanning a large parameter space, where degeneracies between lensing effects and intrinsic source parameters may significantly complicate inference \cite{mishraExploringImpactMicrolensing2024}. A detailed investigation of these degeneracies and the development of robust strategies for parameter estimation are beyond the scope of this paper and are deferred to future research.

\section{\label{sec:LTandspace}Lense-Thirring Precession and parameter space}

This section investigates the effect of LT precession on the geometrical parameters of the lensing system (Sec.~\ref{subsec:LT}), and discusses additional dynamical influences from the AGN environment on the CW source (Sec.~\ref{subsec:DynamicalEffects}).

\subsection{\label{subsec:LT}Lense-Thirring effect}
The spin of a massive central object generates a gravitomagnetic field \cite{2011PhRvD..84l4014N}, which causes a relativistic precession of nearby orbits—known as the LT effect. This is a 1.5 post-Newtonian correction in general relativity, analogous to Larmor precession in a magnetic field. The orbital angular momentum of a test body precesses around the spin axis of the central object due to frame dragging \cite{2017PhRvD..96b3017W,iorioLenseThirringEffectWork2024,fangImpactSpinningSupermassive2019,fangSecularEvolutionCompact2019,Lei2013}.

Following the standard definition in the literature 
\cite{iorioLenseThirringEffectWork2024}, the spin angular momentum of a rotating black hole is given by
\begin{equation}
     \boldsymbol{J} = \boldsymbol{\chi} \frac{G M^2}{c},
\end{equation}
where $\chi \in [-1, 1]$ is the dimensionless spin parameter, with $\chi = 0$ corresponding to a non-rotating Schwarzschild black hole and $\chi = 1$ corresponding to an extremal Kerr black hole. In the case of nearly circular orbits—which is typically valid for CW sources embedded in an accretion disk—the Lense–Thirring (LT) precession of the orbital plane can be described by
\begin{equation}
    \frac{\mathrm{d} \boldsymbol{L}_{\mathrm{orb}}}{\mathrm{d}t} = \boldsymbol{\Omega}_{\rm{LT}} \times \boldsymbol{L}_{\mathrm{orb}},
\end{equation}
where the precession frequency is
\begin{equation}
\boldsymbol{\Omega}_{\rm LT} := \frac{2GJ}{c^{2}a^{3}(1-e^{2})^{3/2}} \, \hat{\chi},
\label{eq:omegaLT}
\end{equation}
with $a$ being the semi-major axis, $e$ the eccentricity, and $\hat{\chi}$ the unit vector along the black hole’s spin axis \cite{fangImpactSpinningSupermassive2019,fangSecularEvolutionCompact2019}.

We define the inclination angle $\iota$ between the observer's line of sight and the orbital angular momentum direction as
\begin{equation}
\cos \iota = \frac{\hat{O} \cdot \boldsymbol{L}_{\mathrm{orb}}}{|\boldsymbol{L}_{\mathrm{orb}}|}.
\end{equation}

In the coordinate system illustrated in Fig.~\ref{fig:example}, the spin axis (aligned with $\boldsymbol{\chi}$) is chosen to define the $z$-axis. Due to the axial symmetry introduced by Lense–Thirring precession, the observer direction $\hat{O}$ can, without loss of generality, be restricted to lie in the $yz$-plane:
\begin{equation}
\hat{O} = (0, \cos\gamma, \sin\gamma),
\end{equation}
where $\gamma$ is the angle between the observer direction and the reference plane.

The unit vector of orbital angular momentum under LT precession is given by
\begin{equation}
\frac{\boldsymbol{L}_{\mathrm{orb}}}{|\boldsymbol{L}_{\mathrm{orb}}|} = (\sin\alpha \cos\Omega, \sin\alpha \sin\Omega,, \cos\alpha),
\end{equation}
where $\alpha$ is the angle between $\boldsymbol{L}_{\mathrm{orb}}$ and the spin axis, and the precession phase evolves as
\begin{equation}
\Omega = \Omega_{\rm LT} t.
\end{equation}

Combining the above, we obtain the inclination angle as a function of time:
\begin{equation}
\cos \iota = \sin\alpha \sin(\Omega_{\rm LT} t) \cos\gamma + \cos\alpha \sin\gamma.
\label{eq:iota}
\end{equation}

\subsection{Dynamical Effects Analysis \label{subsec:DynamicalEffects}}

In this section, we examine other dynamical effects present in the system and their associated timescales, in order to delineate the parameter space relevant to our proposed modulation of the lensing effect by the LT precession. The system under consideration involves a hierarchical triple configuration composed of SMBH-BBH, where various dynamical mechanisms—such as de Sitter-like (dS) precession (1.5PN) \cite{liuBinaryMergersSupermassive2019,yuDirectDeterminationSupermassive2021}, frame dragging to the inner binary (2PN) \cite{fangImpactSpinningSupermassive2019,fangSecularEvolutionCompact2019}, the Lidov-Kozai (LK) mechanism induced by Newtonian tidal interactions \cite{chandramouliKozai-Lidov2022,demeDetectingKozaiLidov2020} and the interaction between disk and CW source \cite{ishibashiEvolutionBinaryBlack2020}. In contrast, for a CW source modeled as a single neutron star, the dynamical effects related to an inner binary orbit can be neglected, resulting in a much simpler scenario. Nonetheless, for the sake of generality, we discuss the dynamical mechanisms of both the inner and outer orbits in this work.

We follow \cite{yuDirectDeterminationSupermassive2021} to examine the relevant dynamical timescales. The characteristic angular frequency of LK oscillations in a hierarchical triple system is given by
\begin{equation}
    \Omega_{\rm LK} = \frac{M_{\rm L}}{(M_1 + M_2)} \left( \frac{a_{\rm i}}{a } \right)^3 \Omega_{\rm i},
    \label{eq:omegaLK}
\end{equation}
where $M_1$ and $M_2$ are the masses of the inner binary components, $a_{\rm i}$ is the semi-major axis of the inner orbit. The term $\Omega_{\rm i}= \sqrt{\frac{G(M_1 + M_2)}{a_{\rm i}^3}}$ denotes the Keplerian angular frequency of the inner binary. The de Sitter (1.5 post-Newtonian) precession angular frequency of the inner orbit is given by
\begin{equation}
    \Omega_{\rm dS} = \frac{3}{2} \frac{G M_{\rm L}}{c^2 a } \, \Omega_{\rm o}.
    \label{eq:omegadS}
\end{equation}
 Moreover, the interaction between the disk and the source only produces significant effects on timescales longer than $10^5$ years \cite{ishibashiEvolutionBinaryBlack2020}; therefore, it can be safely neglected.

\begin{figure}[htbp]
  \centering
  \includegraphics[width=0.45\textwidth]{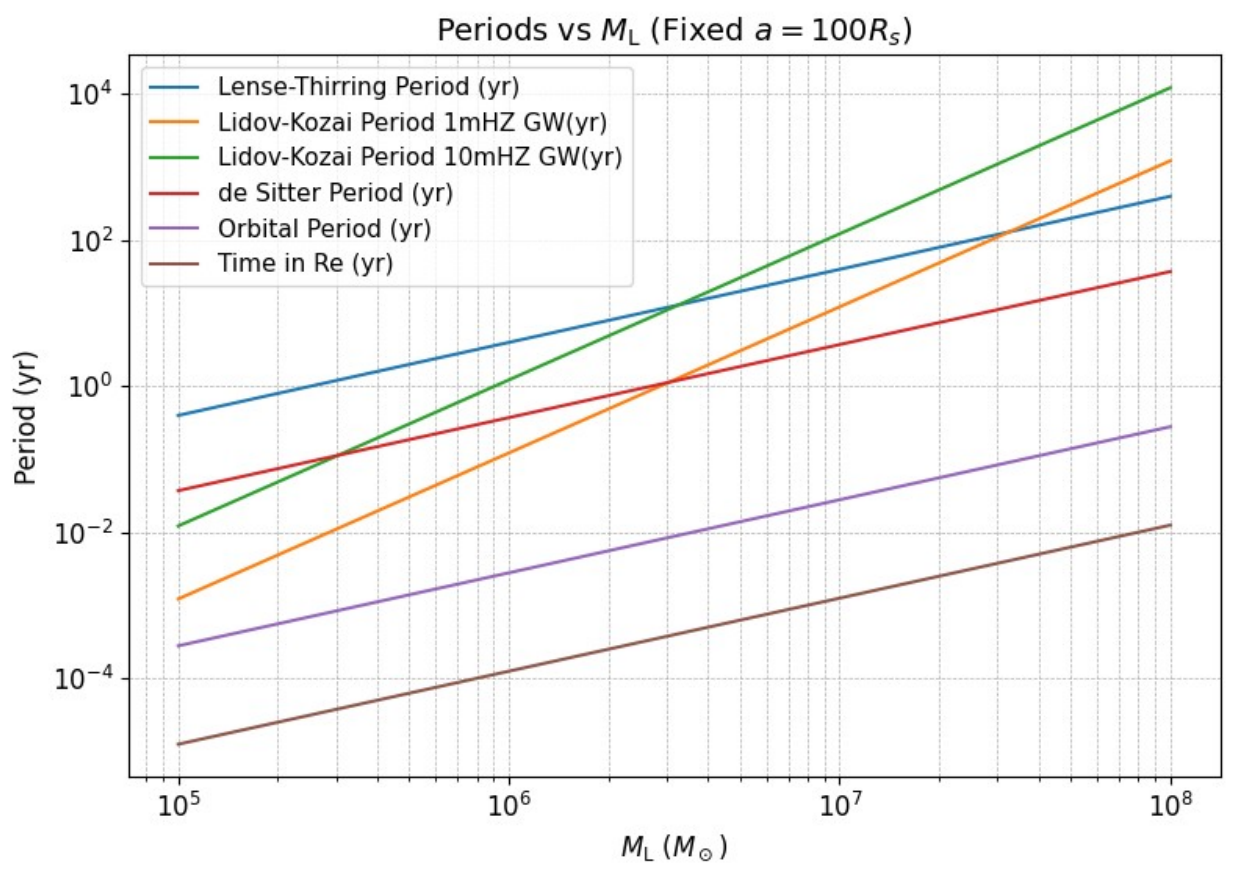}
  \caption{Dynamical timescales as functions of the central SMBH mass $M_{\rm L}$, with the outer orbital separation fixed at $a = 100 R_{\rm S}$. The plotted periods include the Lense-Thirring precession ($P_{\rm LT}$), Lidov-Kozai cycles ($P_{\rm LK}$) assuming inner binaries emitting gravitational waves at 1 mHz and 10 mHz, de Sitter precession ($P_{\rm dS}$), the outer orbital period ($P_{\rm o}$), and Time through $R_{\rm Ein}$. The LT precession timescale is generally much longer than observation timescale. To amplify the subtle LT effect, we rely on the high positional sensitivity provided by lensing observations. Since the LT period increases with the lens mass $M_{\rm L}$, it is necessary to avoid excessively massive lenses to ensure that the LT modulation remains within observationally accessible timescales. Given the short orbital period, the LT-induced precession accumulates over many orbits. Meanwhile, lensing evolves on much shorter timescales, which justifies the neglect of other secular effects when modeling the lensing configuration. }
  \label{fig:period1}
\end{figure}

\begin{figure}[htbp]
  \centering
  \includegraphics[width=0.45\textwidth]{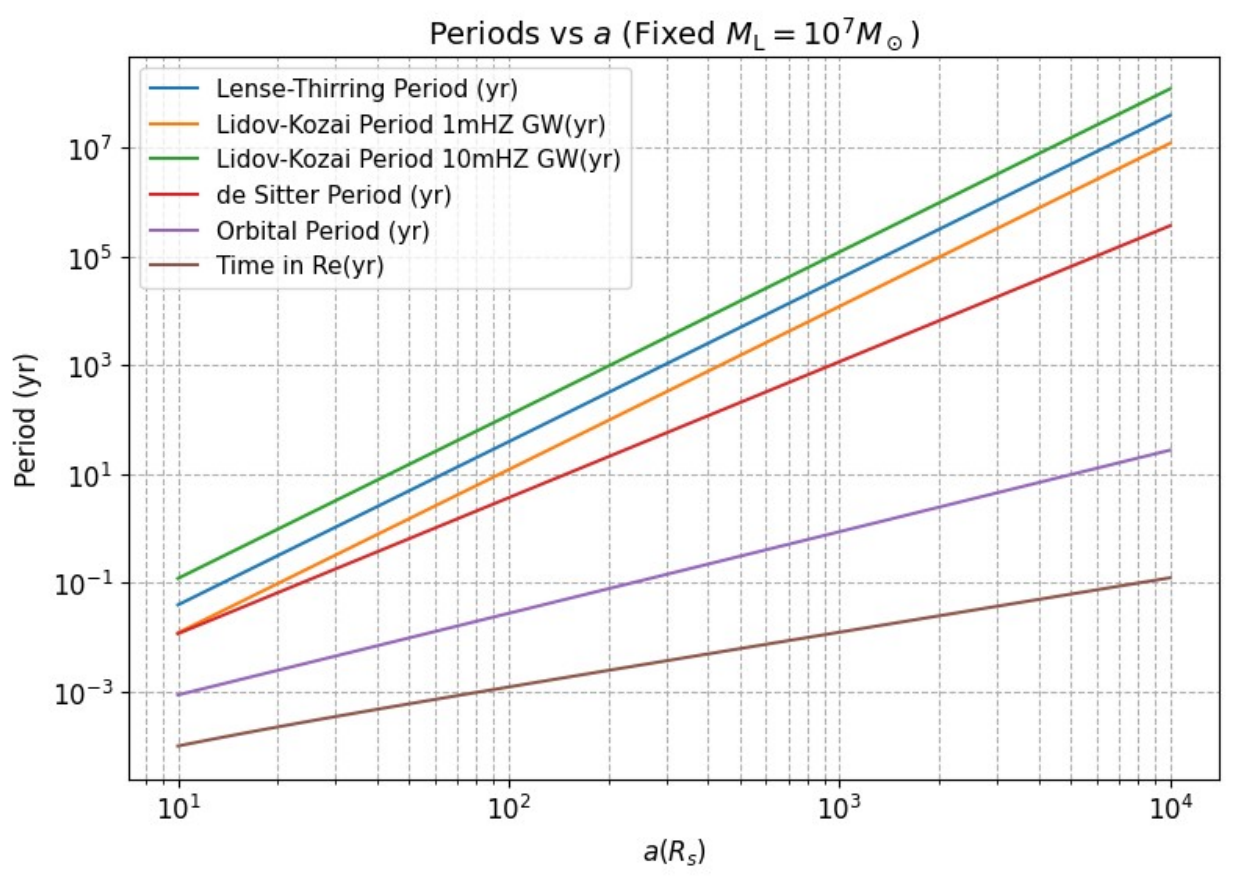}
  \caption{Characteristic timescales as functions of a in units of $R_{\rm S}$, with the SMBH mass fixed at $M_{\rm L} = 10^7 \,M_\odot$. The plotted dynamical timescales are the same as in Fig.~\ref{fig:period1}. The LT precession period decreases rapidly with decreasing orbital radius. This suggests that smaller orbits enhance the effect and improve its detectability.}
  \label{fig:period2}
\end{figure}

By combining Eqs.~\eqref{eq:omegaLT}, \eqref{eq:omegaLK} and \eqref{eq:omegadS}, we obtained the characteristic timescales of different dynamical processes, and we also calculate the timescale for crossing the Einstein radius when $\iota = 90^\circ$ to estimate lensing timescale. Regarding the results of the timescale calculations, we fix $a = 100 R_{\rm S}$ and vary the mass of $M_{\rm L}$, as shown in Fig.~\ref{fig:period1}. Conversely, we fix $M_{\rm L}=10^7\, M_{\odot}$ and vary $a$, as shown in Fig.~\ref{fig:period2}. It can be seen that the LT period lengthens with orbital radius $a$, implying a weaker frame–dragging effect at larger radii, and that the growth of the period with $M_{\rm L}$ is comparatively slower. Therefore, a smaller outer orbital radius can significantly enhance the LT effect. A lower-mass SMBH also contributes to this enhancement; however, reducing the SMBH mass leads to a rapid weakening of the lensing effect, which may render the lens nearly undetectable. For the subsequent discussion, we consider a representative case with $M_{\rm L} = 10^7\,M_\odot$ and $a = 100\,R_{\rm S}$.

For the LK mechanism, during CW emission in the mHz band, the LK timescale is shorter than that of the LT effect. However, as the frequency increases, by around 10 mHz the LK timescale becomes longer than both the LT timescale and the observational window, making its contribution negligible. This is because higher-frequency BBHs are more compact, reducing the effective lever arm for tidal interactions. Moreover, LK oscillations require a minimum mutual inclination between the inner and outer orbits \cite{guptaGravitationalWavesHierarchical2020,migaszewskiNonresonantRelativisticDynamics2011}, while in AGN systems, BBH orbits tend to align with the outer orbit and disk plane \cite{ubachConstrainingEnvironmentCompact2025a,kingAligningSpinningBlack2005}, further limiting the role of LK dynamics.

Although the dS precession and the spin-induced frame dragging by the SMBH act on inner binary have shorter timescales than the LT effect, they do not affect the properties of the outer orbit. This implies that they have no impact on the gravitational lensing effect, which is primarily determined by the outer orbit. Instead, these effects mainly alter the direction of the inner orbital angular momentum, effectively changing the orientation of the BBH system as seen by the observer, while leaving its intrinsic properties unchanged. As a result, they mainly influence the relative amplitude and polarization modes of the gravitational wave signal, with negligible impact on its frequency content. Moreover, the dynamical timescales associated with these effects are much longer than the time in Einstein radius ($R_{\rm Ein}$), which produces short, sharply periodic pulses. Within the parameter space we consider, such lensing features are qualitatively distinct from the signatures of the aforementioned dynamical effects and can be clearly distinguished. For example, \cite{yuDetectingGravitationalLensing2021} illustrated the waveforms under similar sets of parameters, highlighting the two effects is readily distinguishable.

\section{LT Effects on lensing \label{sec:LTonlense}}
In this section, we investigate the influence of LT precession on the lensing effect. Variations in the Einstein radius crossing time and peak magnification are presented in Sec.~\ref{subsec:LensingFeatures}. In Sec.~\ref{subsec:Matched-Filter}, the strength and detectability of the lensing effect are quantified through matched filter analysis. Sec.~\ref{subsec:probability} explores the increased probability of lensing event occurrence.

\subsection{Lensing Features\label{subsec:LensingFeatures}}

\begin{figure}[htbp]
  \centering
  \includegraphics[width=0.45\textwidth]{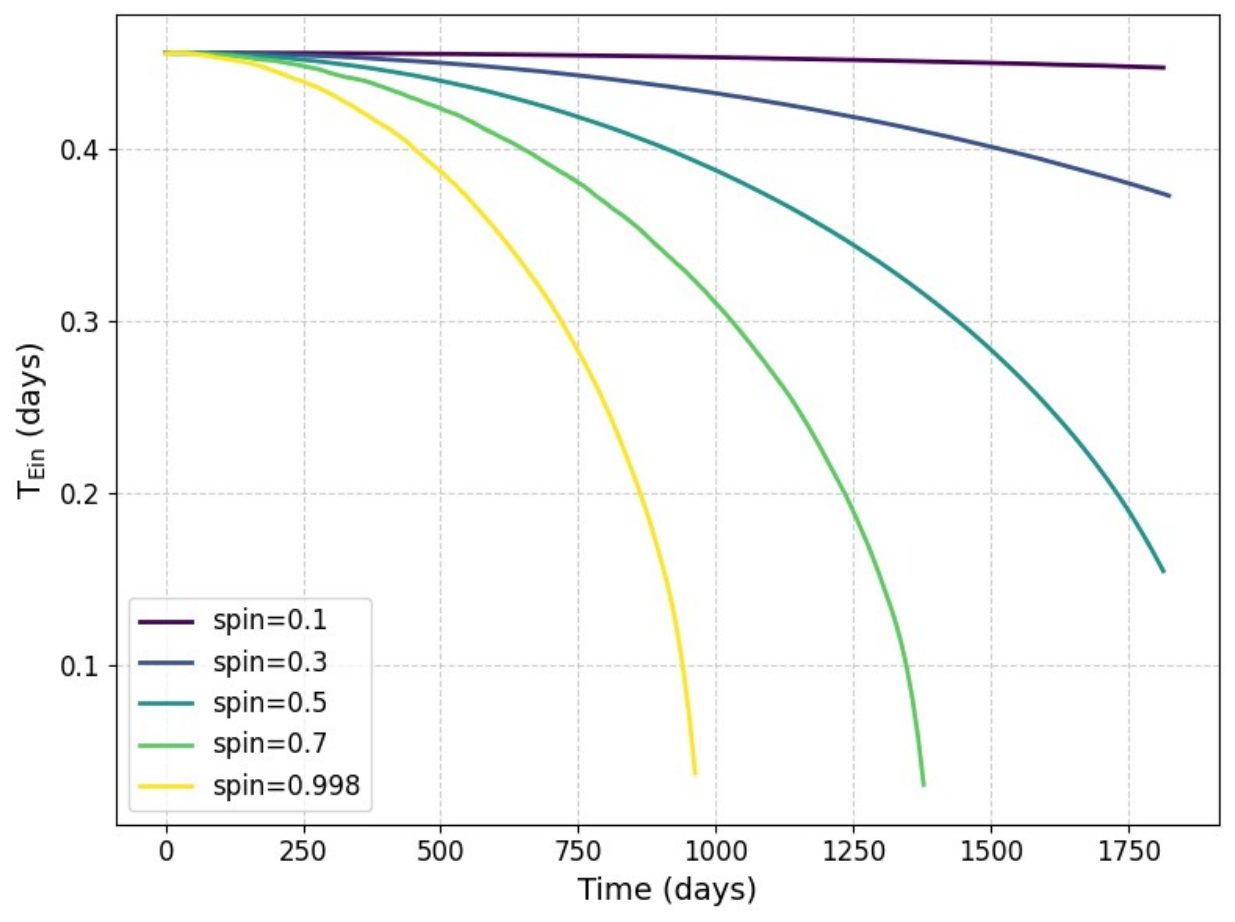}
  \caption{The evolution of the lensing duration $T_{\rm Ein}$, defined as the time interval during which the source passes within the Einstein radius for each orbital pass separately, is shown as a function of the system’s evolution time (horizontal axis). The lensing duration decreases as the system deviates from an edge-on orientation, with more rapid shortening observed for higher spin values. For low-spin cases, the duration remains nearly constant.}
  \label{fig:Techange}
\end{figure}

\begin{figure}[ht]
  \centering
  \includegraphics[width=0.45\textwidth]{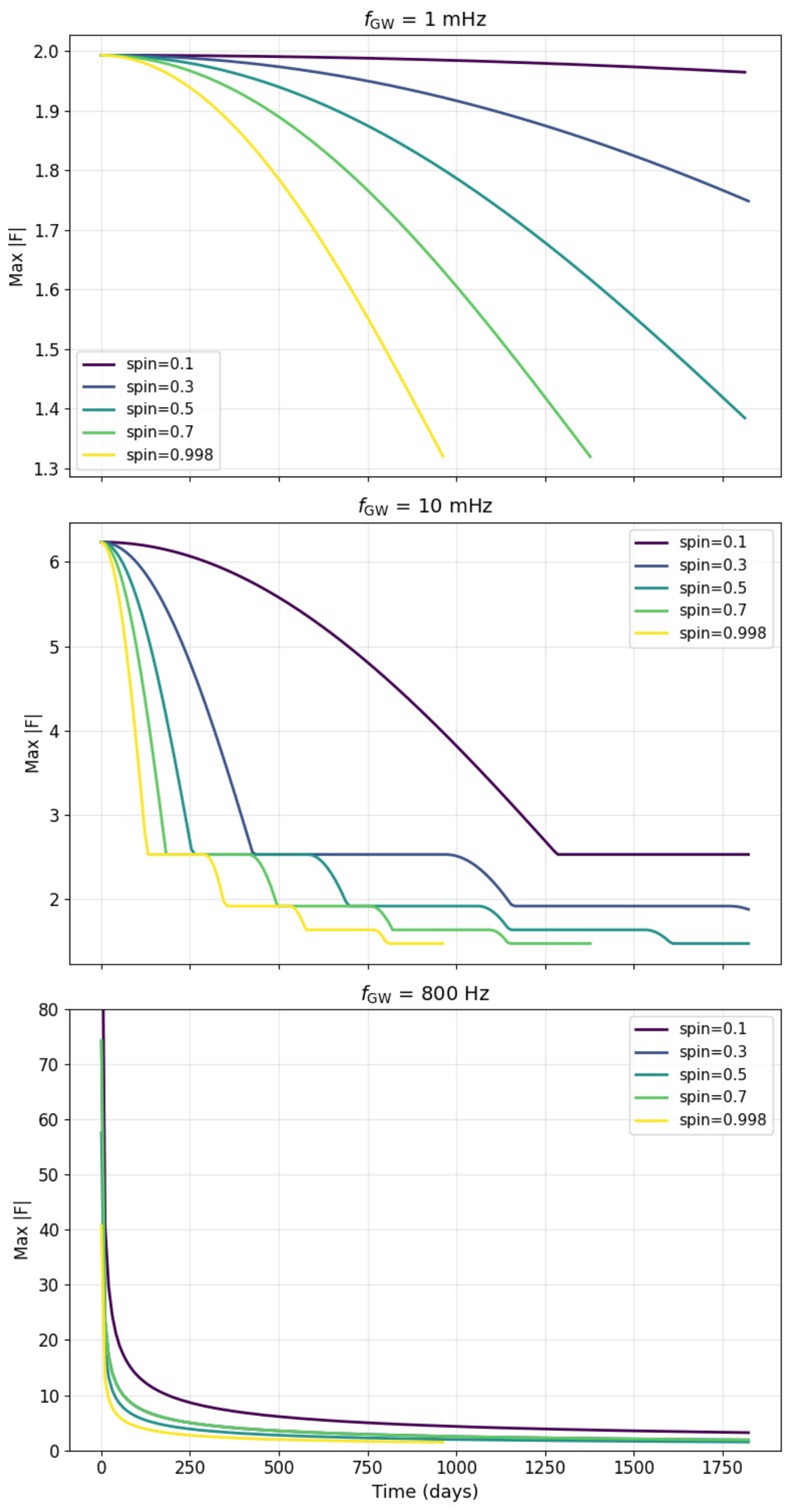}
  \caption{Maximum values of $\left| F \right|$ within each interval of repeated lensing events, plotted as a function of time. The three subpanels correspond to different frequency bands. Peak values of $\left| F \right|$ decrease monotonically over time, with faster decay observed for higher spin parameters. The decay rate is slower at lower frequencies, while higher frequencies exhibit a rapid initial decline. This behavior reflects the increasing sensitivity of $F$ to the source position near $\eta = 0$ upon entering the geometrical optics regime.}
  \label{fig:MAXF}
\end{figure}

In this section, we investigate the impact of the LT effect on gravitational lensing. Our analysis focuses on several key aspects, including the time at which the source crosses the Einstein radius ($T_{\rm Ein}$), the maximum amplification of the waveform amplitude. Since repeated lensing is considered, the effect of each individual lensing event is quantified, and the time of each orbital phase $\phi = 0$—corresponding to the center of the lensing effect—is used to trace the evolutionary timescale of the system. In Figs.~\ref{fig:Techange} to \ref{fig:mismatch3}, the x-axis [Time (days)] represents the evolutionary time of the system, measured from the onset of the strongest lensing configuration, with each data point corresponding to a successive orbital passage. The curves in these figures are formed by connecting discrete data points, each corresponding to an individual lensing event. The y-axis depicts the physical conditions during each lensing event, while the x-axis indicates the specific evolutionary time at which that event occurred.

We consider the following parameter set as an example case: $M_{\rm L} = 10^7\,M_\odot$ and $a = 100R_{\rm S}$. Under these conditions, the orbital period is approximately $T_{\rm o} \approx 10.13$ days, and the LT precession period is about 40 years for a dimensionless spin parameter $\chi = 0.94$ (similar to that of Sgr A*). Our analysis begins from the edge-on configuration ($\iota = 90^\circ$), corresponding to the case where the lensing effect is strongest and we follow its evolution over a 5-year observation period. We examine three representative gravitational wave frequencies: $f_{\rm GW} = 10^{-3}$ Hz, $10^{-2}$ Hz (BBH), and 800 Hz (NS or boson clouds). In the following analysis, the transmission factor is computed using the full wave optics formalism for $10^{-3}$ Hz and $10^{-2}$ Hz, while the geometrical optics approximation is employed for 800 Hz. The dimensionless spin parameter of the central object is varied across five values: $\chi = 0.1$, $0.3$, $0.5$, $0.7$, and $0.998$ \cite{thorneDiskAccretion1974}. When analyzing the impact of spin, we fix the geometric parameters to $\alpha = 20^\circ$ and $\gamma = 0^\circ$. We also briefly explore the role of these geometric parameters. While $\alpha$ and $\gamma$ determine the amplitude and central value of $\iota$ variation, the spin parameter $\chi$ predominantly controls the frequency of this variation through the LT precession.

As shown in Fig.~\ref{fig:Techange}, the time during which the source passes through the Einstein radius decreases as the orbital inclination of the system deviates from the edge-on configuration during its evolution. This indicates that the lensing duration is progressively shortening. The rate of this decrease is more pronounced for systems with higher spin, while for low-spin cases (e.g., $\chi = 0.1$), the duration remains nearly constant.

It should be noted that this lensing duration may serve as an observable quantity. Since the CW frequency can be directly extracted from the waveform, the corresponding transmission factor for a given frequency—specifically, the value associated with $\eta = 1$—can be determined (see Eq. \eqref{eq:transmission_factor}). This transmission factor can be inferred from both the waveform (see Eq. \eqref{eq:lensedCW}) and the SNR \cite{suyamprakasamMicrolensingLongdurationGravitational2025}. The lensing duration, denoted as $T_{\rm Ein}$, can be characterized by the time interval during which the waveform amplitude exceeds the amplification threshold($F(f, \eta =1) $).

We proceed to evaluate the maximum value of $\left| F \right|$ within each interval of repeated lensing events. Notably, this quantity can be directly inferred from the waveform. As illustrated in Fig.~\ref{fig:MAXF}, the peak values of $\left| F \right|$ exhibit a monotonic decrease over time, with the decay proceeding more rapidly for higher spin values.

A comparison across the three subpanels reveals that the overall decay rate is slower at lower frequencies, whereas at higher frequencies, a pronounced early-time decay is observed. This trend can be attributed to the transition into the geometrical optics regime, where $F$ becomes highly sensitive to the source position near $\eta = 0$. In contrast, the contribution from the wave optics regime remains relatively weak, as previously demonstrated in Figs. \ref{fig:TF001} and \ref{fig:TF800}. 

Interestingly, for the 10 mHz GW signal (middle panel), we first noticed that the maximum value of the transmission factor $F$ exhibits a plateau over a finite range. To investigate this behavior, we further analyzed the evolution of $F$ as a function of the inclination angle $\iota$. As the system gradually deviates from the edge-on configuration, we observe that the central peak of $F$ initially decreases from its maximum value toward zero. Subsequently, the two side peaks begin to merge toward the center, forming a new dominant peak. During the merging process, the maximum value of $F$ remains constant for a certain interval, giving rise to the observed plateau. After the merger completes, the new peak then begins to decay. This sequence of changes closely resembles the contraction and disappearance of central interference fringes in optics. In both the 1mHz and 800Hz cases, no plateau-like feature in $F_{\text{max}}$ is observed. In the 1mHz case, only a single peak lies within the Einstein radius, and the peak value of the transmission factor $F$ decreases steadily. For 800Hz, the oscillations are densely packed, effectively washing out any sustained maximum.

\subsection{Matched-Filter Analysis and Effect Detectability \label{subsec:Matched-Filter}}
In the subsequent analysis, we employ matched filter analysis to quantitatively assess the strength of the lensing effect \cite{yuDetectingGravitationalLensing2021,chanDetectabilityLensedGravitational2024,ubachConstrainingEnvironmentCompact2025a}. Specifically, we compute the mismatch between the lensed and unlensed waveforms, where a larger mismatch indicates a more pronounced impact of lensing \cite{bondarescuCompactBinariesLens2023}. The match between the lensed and unlensed waveforms is defined as \cite{cutlerGravitationalWavesMerging1994}
\begin{align}
\mathcal{M}(h_{\rm UL}, h_{\rm L}) = \max _{\varphi_{0}, t_{0}}\frac{\langle h_{\rm L} | h_{\rm UL} \rangle}{\sqrt{\langle h_{\rm L} | h_{\rm L} \rangle \langle h_{\rm UL} | h_{\rm UL} \rangle}},
\label{eq:match}
\end{align}
which is optimized over the GW phase $\varphi_0$ and the time of arrival $t_0$, and the noise-weighted inner product is given by
\begin{equation}
\langle h_{\rm L} | h_{\rm UL} \rangle \equiv 2 \int \frac{\tilde{h}^\ast_{\rm L}(f) \tilde{h}_{\rm UL}(f) + \tilde{h}_{\rm L}(f) \tilde{h}^\ast_{\rm UL}(f)}{S_n(f)}\, df,
\label{eq:inner_product}
\end{equation}
with $S_n(f)$ denoting the one-sided noise power spectral density (PSD). The mismatch is then defined as
\begin{equation}
\mathcal{MM} = 1 - \mathcal{M}.
\end{equation}
To simplify the computational procedure, the detector response functions were not incorporated in the calculation of the mismatches. If the mismatch $\mathcal{MM}$ exceeds the threshold $\mathcal{MM}_{\rm th} \simeq \frac{1}{2\rho^2}$, it implies that the systematic deviation caused by the intrinsic differences between the two waveforms dominates over the statistical uncertainty introduced by noise fluctuations. In such cases, the two waveforms are deemed distinguishable which means the effect is in principle detectable \cite{cutlerGravitational1994,lindblomModelWaveformAccuracy2008a,yuDetectingGravitationalLensing2021,aliDetectabilityStronglyLensed2023,thompsonUse2025}. However, this conclusion does not consider possible parameter degeneracies and other effects that may contribute to the observed mismatch. In practice, a realistic and robust assessment of detectability requires more extensive modeling over the full parameter space. As part of a preliminary conceptual discussion, we employ the mismatch calculation to simplify the problem.

We evaluate the waveform mismatch using the \texttt{PyCBC} package \cite{usmanPyCBC2016}, For the PSD of the ground-based detectors, we adopt the \texttt{aLIGOAdVO4IntermediateT1800545} \cite{abbottProspects2020} and employ \texttt{analytical\_psd\_lisa\_tdi\_AE\_confusion} for the expected performance of the space-based  detector LISA \cite{LISA_LDC_Manual_2020}, considering two representative SNR values, namely $\rho = 4$ and $8$ for each individual “lensing event”.

\begin{figure}[htbp]
  \centering
  \includegraphics[width=0.45\textwidth]{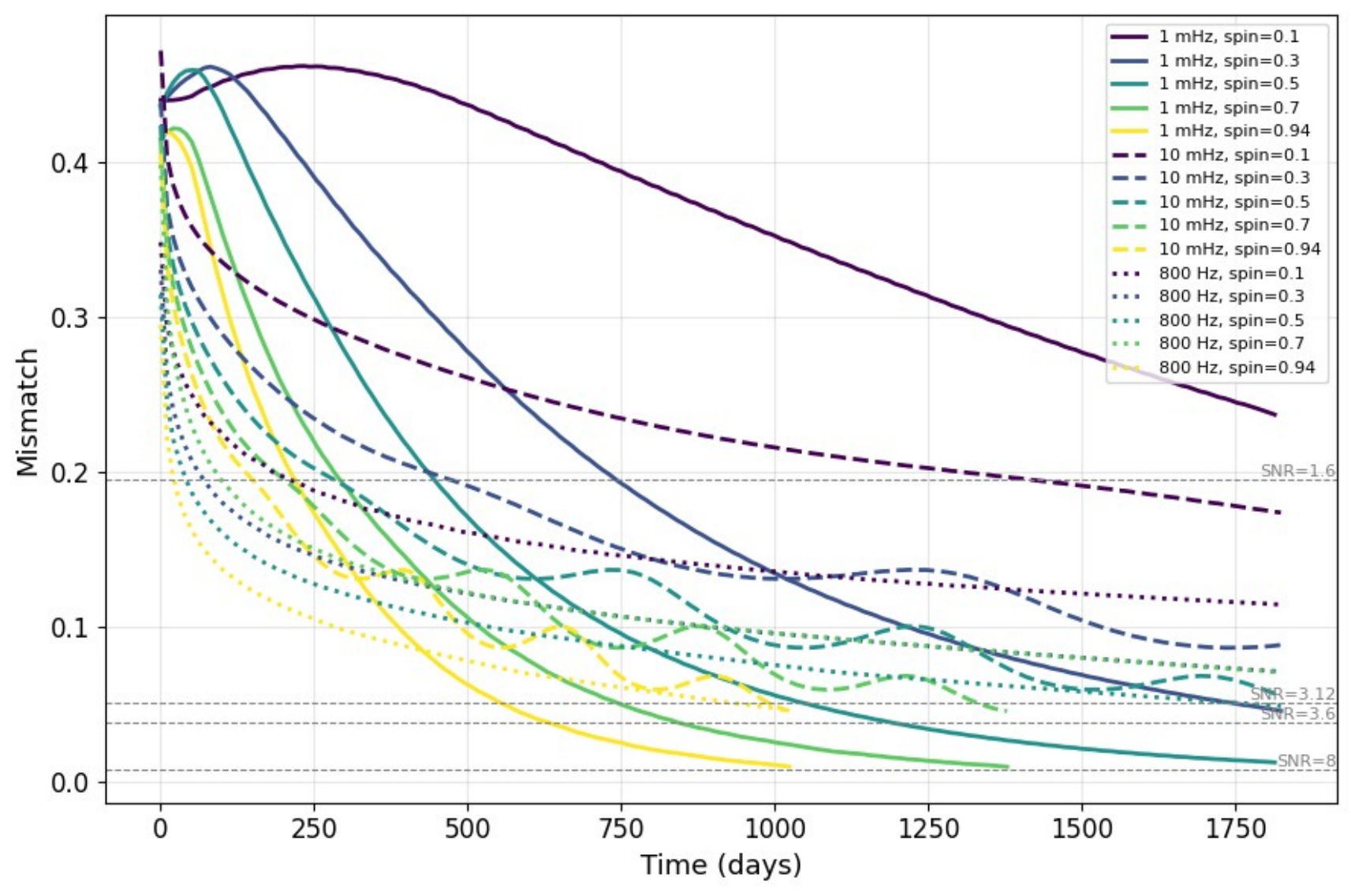}
  \caption{Mismatch between lensed and unlensed waveforms computed at each lensing event. Time intervals are same as those in Fig.~\ref{fig:TF001}. Different line styles correspond to frequencies of 1 mHz (solid), 10 mHz (dashed), and 800 Hz (dotted). Colors indicate spin values. Gray dashed lines mark mismatch thresholds for different SNRs.}
  \label{fig:mismatch1}
\end{figure}

In Fig.~\ref{fig:mismatch1}, we compute the mismatch between the lensed and unlensed waveforms at each lensing event. The timing of each mismatch measurement is aligned with the choice of $T_{\rm Ein}$ in Fig.~\ref{fig:TF001}, with the horizontal axis corresponding to the evolution time defined by the moment of each lense event when the source phase $\phi = 0$. In the plot, the frequencies 1 mHz, 10 mHz, and 800 Hz are represented by solid, dashed, and dotted lines, respectively, while the color encodes the spin magnitude, following the same convention used in previous figures. The gray dashed lines indicate the mismatch threshold for distinguishability, based on the corresponding SNR.

For a fixed frequency, we observe that the mismatch decreases more rapidly for higher spin values, reflecting the stronger lensing effects induced by faster spins. On the other hand, for a fixed spin, the mismatch evolution depends sensitively on the GW frequency: at lower frequencies, the mismatch initially decreases more slowly but later shows a faster decline; in contrast, higher-frequency CWs exhibit a rapid initial drop followed by a more gradual decline. This behavior can be attributed to the lensing transmission factor $F$ in the geometrical optics regime, which becomes highly sensitive to the source position near $\eta = 0$, making the waveform more sensitive to geometric configuration at early times. However, as $|F|$ remains relatively large in the geometrical optics regime, the lensing effect weakens more slowly in later stages, resulting in a slower mismatch decay. Within this observational timescale, the lensing effect is predominantly observable. The only exceptions occur for systems with high spin values and low gravitational wave frequencies $f_{\rm GW}$, where the lensing effect may become undetectable.

Additionally, we observe oscillatory features superimposed on the overall mismatch decay at 10 mHz, similar to the features displayed in the middle panel of Fig.~\ref{fig:MAXF}. This oscillatory pattern can be interpreted as resulting from the behavior of the transmission factor: when the two side peaks begin to merge toward the center, the peak value of $F$ remains nearly constant. Consequently, the mismatch exhibits little variation during this stage. However, as the peak value of $F$ starts to decrease, the mismatch drops rapidly, giving rise to the observed oscillations.

Starting from a representative source with SNR of $8$ under unit detector response, we incorporate the effect of the detector response through the factor $R$ appearing in Eq.~\eqref{eq:SNR}. Specifically, we consider four cases: unit response ($R=1$), the sky- and polarization-averaged response for ground-based detectors such as LIGO ($R=0.45$), the corresponding averaged response for LISA ($R=0.39$), and a conservative estimate $R=0.2$. The resulting effective signal-to-noise ratios are therefore
\[
\rho = 8,\; 3.6,\; 3.12,\; \text{and } 1.6 ,
\]
as shown in the figure.

As illustrated, once realistic detector response effects are taken into account, the effective SNR is reduced compared to the idealized unit-response case. This reduction leads to an increase in the mismatch threshold required to distinguish lensed signals, implying that stronger lensing-induced waveform differences (i.e., larger mismatches) are necessary for the lensing effect to be identifiable. Consequently, the detectability criterion becomes more stringent when realistic detector responses are considered. The impact of this effect is found to be qualitatively the same in the subsequent analyses.

\begin{figure}[htbp]
  \centering
  \includegraphics[width=0.45\textwidth]{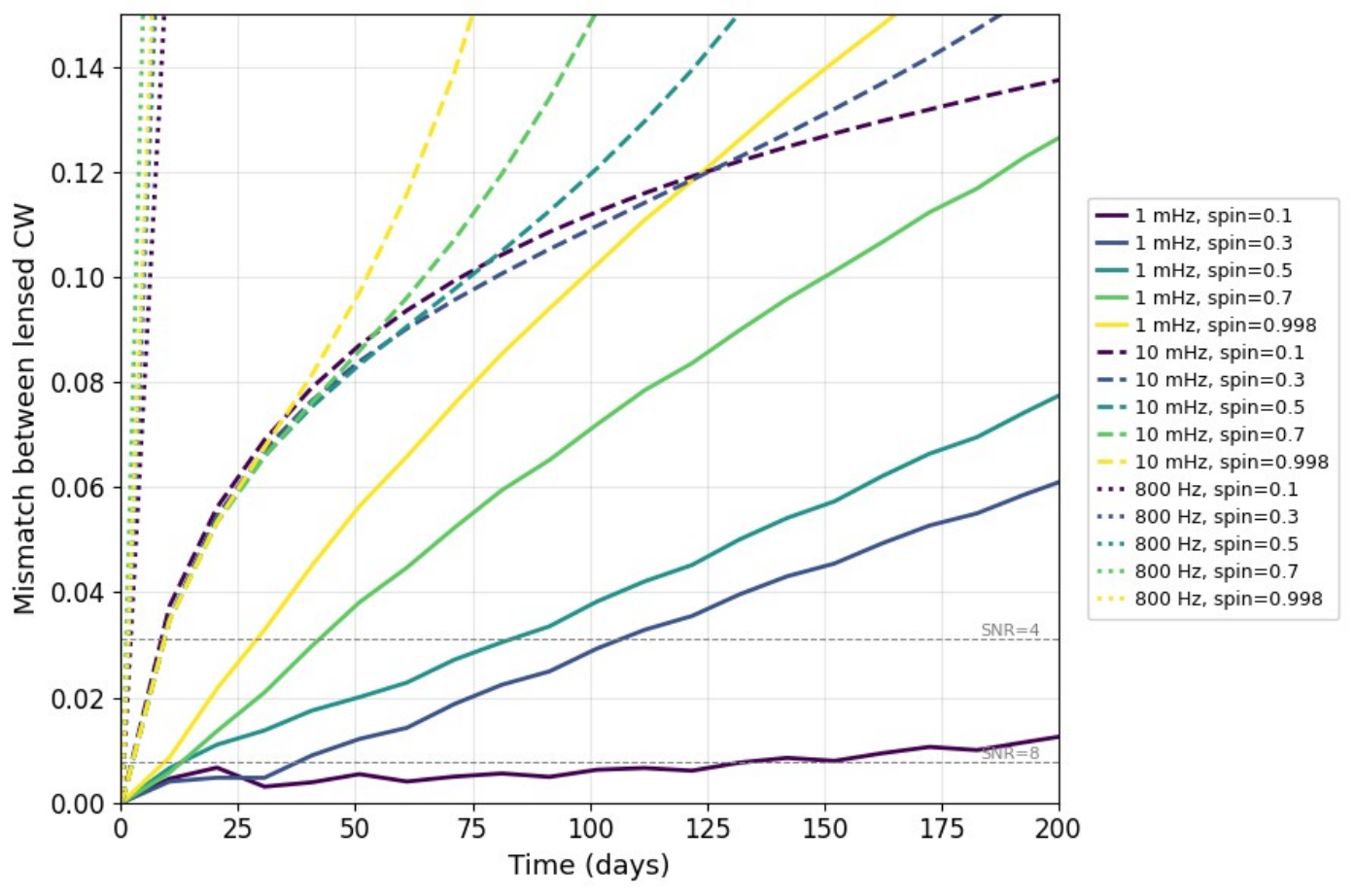}
  \caption{Mismatch between the first lensing waveform and subsequent lensed waveforms. Time intervals and legend conventions are consistent with Fig.~\ref{fig:mismatch1}. The plot shows how mismatch evolves with time for different frequencies and spins, indicating when LT precession effects become noticeable.}
  \label{fig:mismatch2}
\end{figure}

Additionally, we computed the mismatch between lensed waveforms, specifically comparing the first (edge on) lensed waveform with subsequent lensed waveforms. The results are presented in Fig.~\ref{fig:mismatch2}, where the time intervals and legend conventions for the mismatch calculation are consistent with those used in Fig.~\ref{fig:mismatch1}. This analysis effectively addresses the ideal observational timescale required to detect LT effects from lensing. For instance, for a CW signal at 1 mHz with SNR of 8 and spin parameter $\chi = 0.1$, approximately 130 days of observation are necessary to discern the lensing waveform variation induced by LT precession. In the absence of LT effects, the lensed waveform remains identical across repeated lensing events, resulting in a mismatch of zero. Consequently, for the low $f_{\rm GW}$ case, higher spins correspond to shorter required observation times to detect LT effects. Moreover, for the early-time oscillatory behavior, by plotting the waveform and the transmission factor $F$, we observe that near the well-aligned region, the phases of the waveform and $F$ can undergo abrupt jumps, which in turn leads to instabilities in the mismatch computation. In contrast, for higher frequency bands, the substantial difference in the transmission factor $F$ enables us to observe significant discrepancies already at the second occurrence of lensing, thus one orbital period is sufficient.

In addition, we investigate the influence of the geometric parameters $\alpha$ and $\gamma$. As shown analytically in Eq.~\eqref{eq:iota}, both parameters contribute to the central value and amplitude of the inclination angle $\iota$, but do not affect its modulation period. Their effects are thus relatively straightforward to interpret. In what follows, we focus on $\alpha$ as a representative example adopting the same parameters as before and fixing the $f_{\rm gw}=1$ mHz.

\begin{figure}[htbp]
  \centering
 \includegraphics[width=0.45\textwidth]{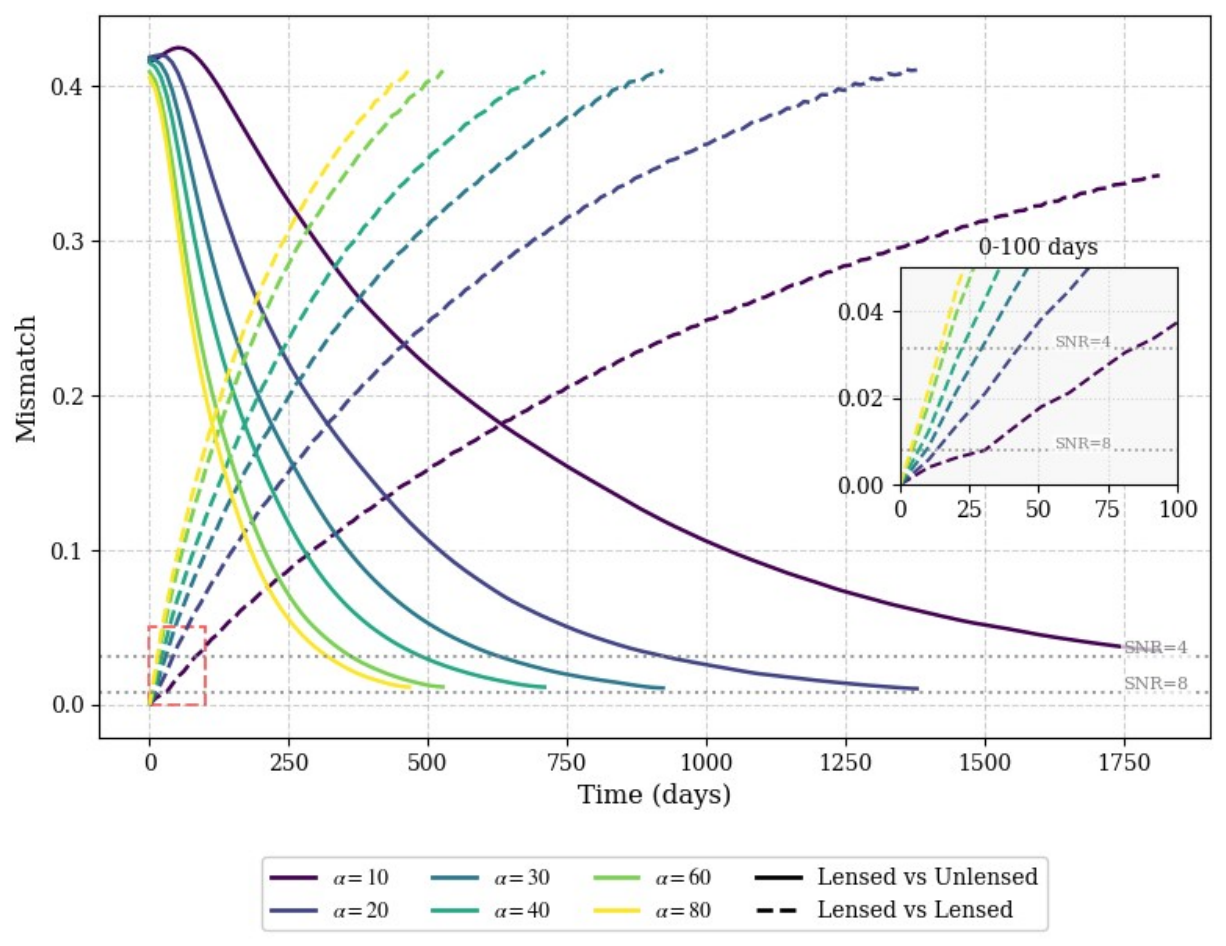}
  \caption{Mismatch comparison illustrating the effect of varying the geometric parameter $\alpha$ under fixed spin parameters $\chi=0.7$. Solid lines represent mismatches between lensed and unlensed waveforms of each lensing event, while dashed lines show mismatches between the first lensed waveform and subsequent lensed waveforms. The right panel shows a magnified view of the region enclosed by the red dashed box. As $\alpha$ increases, the influence of LT precession on the waveform becomes more pronounced, resulting in stronger modulation and more rapid variations in the mismatch. The effect is observable within the observational timescale of 100 days.}
  \label{fig:mismatch3}
\end{figure}

Fig.~\ref{fig:mismatch3} presents the mismatch between lensed and unlensed waveforms (solid lines), following the same convention as in Fig.~\ref{fig:mismatch1}, and the mismatch between lensed waveforms with different configurations (dashed lines), analogous to Fig.~\ref{fig:mismatch2}. We examine the impact of varying $\alpha$ under identical spin parameters $\chi=0.7$. It is evident that larger values of $\alpha$ enhance the influence of LT precession on the lensing signatures. Consequently, the LT-induced modulation manifests on shorter timescales as $\alpha$ increases.

While the variation in $\alpha$ may introduce degeneracies with spin effects in very short-duration observations, for the representative system considered here and an observation time of five years, the distinction is clear. The mismatch as a function of $T_{\rm Ein}$ essentially traces the modulation of $\iota$. Since the geometric angles do not alter the period of $\iota$, the effects of $\alpha$ and spin can, in principle, be disentangled through parameter estimation.

\subsection{\label{subsec:probability}Increase of Probability}
We compute the probability of a lensing event within the system by using the Einstein radius as the criterion \citep{dorazio_periodic_2018,leong_constraining_2025}. In previous work discussing the repeating lens regime without considering spin, the outer orbit was approximated as a ring randomly oriented with respect to the line of sight \citep{dorazio_repeated_2020}. The probability of a significant lensing event is then given by the likelihood that this ring falls within the Einstein radius of the lens. In this section, the assumption of initial alignment is dropped to facilitate a comprehensive probabilistic estimation.

Following \cite{dorazio_repeated_2020,leong_constraining_2025}, the Einstein radius criterion
\begin{equation}
\beta \leq \theta_{\rm{Ein}}
\end{equation}
can be translated into a constraint on the inclination angle $\iota$ in a repeated lensing system:
\begin{equation}
\label{eq:iotacri}
\sin\iota > x = \frac{\sqrt{1 + {(\frac{a}{R_{\rm S}}})^2} - 1}{{\frac{a}{R_{\rm S}}}}
\end{equation}

Therefore, assuming that the inclination angle $\iota$ is initially uniformly distributed in the range $[0, \pi]$, the probability of repeated lensing for a non-spinning system is given by
\begin{equation}
P_{\text{\rm{ns}}}(\sin\iota > x) = 1 - \frac{2}{\pi} \arcsin(x).
\end{equation}

We aim to determine how much the probability increases under the spinning condition:
if the condition $\sin\iota > x$ is satisfied at any point during the observation period $T_{ob}$, the event is considered to have occurred.

The evolve of $\iota(t)$ is 
\begin{equation}
    \cos\iota(t) = \sin\alpha \sin(\Omega_{\rm LT} t) \cos\gamma + \cos\alpha \sin\gamma = m \sin\Omega + n
\end{equation}
where
$$
\Omega = \Omega_{\rm LT} t,\quad
m = \sin\alpha \cos\gamma,\quad n = \cos\alpha \sin\gamma.
$$

This means that the variation of $\iota$ follows a sine-like function, so it is unreasonable to still use a uniform distribution of $\iota$ for the estimation. What we need to estimate is the probability of the system being lensed given $\alpha$ and $\gamma$, and on that basis, consider the increase in probability due to spin.
We require:
$$
\sin\iota(t) > x \quad \Leftrightarrow \quad |m\sin\Omega + n| < \sqrt{1 - x^2} \equiv C.
$$
For the case where the SMBH has no spin, $\Omega$ is a constant, and therefore the probability is:
\begin{equation}
P_{\text{ns}}(\sin\iota > x|\alpha,\gamma)=  
\frac{1}{\pi} \left( \arcsin{\frac{C - n}{m}}  - \arcsin {\frac{-C - n}{m}}  \right).
\label{eq:Pns}
\end{equation}

Consider $\Omega$ evolves over time due to Lense-Thirring precession. Over an observation time $T_{\rm ob}$, the total phase range covered is:

\begin{equation}
\Delta \Omega = \Omega_{\rm LT} T_{\rm ob}.    
\end{equation}
Suppose that the event of lensing can occur only when satisfy (\ref{eq:iotacri}), which corresponds to a favorable phase range on the circle. $P_{\text{ns}}$ in Eq. \eqref{eq:Pns} denotes the probability of randomly finding the system in such a favorable configuration for a given $\alpha$ and $\gamma$. This corresponds to a total "good" phase arc length of $2\pi P_{\text{ns}}$, while the complementary "bad" phase region spans an arc length of $2\pi (1 - P_{\text{ns}})$.

The probability that the event is missed entirely during the observation—i.e., the entire precession interval lies within the unfavorable region—is:

\begin{equation}
    P_{\text{miss}} =
\begin{cases}
\frac{2\pi (1 - P_{\text{ns}}) - \Omega_{\rm LT} T_{\rm ob}}{2\pi}, & \text{if } \Omega_{\rm LT} T_{\rm ob} < 2\pi (1 - P_{\text{ns}}), \\
0, & \text{otherwise}.
\end{cases}
\end{equation}
Therefore, the probability that the event occurs at least once during the observation is:
\begin{multline}
P_{\text{obs}} = 1 - P_{\text{miss}} = \\
\begin{cases}
P_{\text{ns}} + \dfrac{\Omega_{\rm LT} T_{\rm ob}}{2\pi}, & 
\Omega_{\rm LT} T_{\rm ob} < 2\pi (1 - P_{\text{ns}}), \\
1, & \text{otherwise}.
\end{cases}
\end{multline}
Hence, the increase in probability due to finite observation time is:
\begin{equation}
 \Delta P |_{\alpha,\gamma} = P_{\text{obs}} - P_{\text{ns}} = \min\left( \frac{\Omega_{\rm LT} T_{\rm ob}}{2\pi},\ 1 - P_{\text{ns}} \right).   
\end{equation}
In the system configuration considered above with 5 years observation, the relative variation $\Delta P |_{\alpha,\gamma}$, evaluated at fixed geometric angles $\alpha$ and $\gamma$, is approximately 1\% for a spin parameter of $\chi = 0.1$, and increases to about 13\% for $\chi = 0.998$, assuming the same system parameters as before, with $M_{\rm L} = 10^7\, M_\odot$ and $a = 100 R_{\rm S}$.

\section{Conclusion and Discussion\label{sec:discussion}}

In this work, we investigate how LT precession, caused by the spin of a SMBH, modulates the repeated gravitational lensing of CW sources embedded in AGN disks. We demonstrate that the LT-induced precession alters the inclination angle $\iota$ between the source orbital angular momentum and the observer over time, thereby modulating the duration and magnification of repeated lensing events. Using the matched-filter analysis, we show that these modulations leave observable imprints in the waveform—especially for high-spin SMBHs, smaller orbital radii $a$, high-frequency and high-amplitude CW signals, and large misalignment angles $\alpha$. In addition, we find that LT precession not only modulates waveform features but also increases the overall probability of detecting lensing events within finite observation times. This modulation effect may also be applicable to lensing of continuous EM sources.

Although we have used $M_{\rm L} = 10^7 \,M_\odot$ and $a = 100 R_{\rm S}$ as illustrative examples, similar effects persist for other parameter choices. The LT effect becomes more prominent only when $a$ is relatively small. However, if the source is too close to the lens, it may reside entirely within the strong-field region, in which case more detailed numerical computations are required. It is worth emphasizing that the parameters $M_L$ and $a$ are in principle measurable. By extracting the line-of-sight orbital velocity from the Doppler effect and determining the repeating lensing period, one can constrain these parameters using Kepler’s third law. For sources that are only visible during lensing events, the Doppler shift along the line of sight may not be directly observable. Nevertheless, due to the nature of lensing in the geometrical optics regime, GW signals from different directions can become accessible, allowing the measurement of transverse velocities—and thus enabling a reconstruction of the source’s motion \cite{savastanoLensSgrIdentifying2024,samsingConstrainingProperMotion2025}.

The previous sections addressed spin-induced modifications of the source orbit and their indirect effect on lensing. In contrast, we here briefly discuss the direct influence of SMBH spin on the lensing as well as the transmission factor.
We employ the point-mass model to compute the transmission factor induced by the SMBH acting as a lens. This result is obtained under the assumption of a Schwarzschild black hole \cite{Takahashi_2003}. For the Kerr case, \cite{kubotaSpinWaveOptics2024} has computed the transmission factor in the low-frequency regime within wave optics, showing that the spin of the black hole introduces only a small correction to the non-spinning result—an effect that is challenging to detect in practical observations. Another study \cite{baraldoGravitationally1999,bongaWaveOpticsRotating2025}, under the slow-spin approximation, found that the leading-order contribution effectively results in a global shift of the source position, which is equivalent to a coordinate translation and thus does not affect our calculations. However, in the general Kerr spacetime, gravitational lensing behavior becomes significantly more complex, and a complete expression for the transmission factor is still lacking. Recent theoretical progress in this area can be found in \cite{serenoAnalytical2006,jamesGravitational2015,grallaLensing2020,frostGravitational2024}. Moreover, the modulation of the source position by black hole spin, as considered in this work, warrants further investigation, as the lensing effect of spin is instantaneous, while its influence on the source trajectory can accumulate over time.

Our work reveals intriguing modulational effects induced by SMBH spin in lensed gravitational waveforms. If such repeatedly lensed CWs can be observed in the future, the proposed model might be used to interpret some of these modulational signatures and to constrain the physical parameters of the lensing system. While both the geometric inclination and the spin magnitude influence the rate at which the source orbit shifts relative to the Einstein radius, as discussed in Section~\ref{subsec:Matched-Filter}, the expression for the inclination angle $\iota$ (Eq.~\eqref{eq:iota}) indicates that the frequency of variation is modulated solely by the spin. This implies that, in principle, the spin may be constrained through parameter fitting. Nonetheless, a full Bayesian inference analysis would be required for rigorous and statistically robust parameter estimation.

\begin{acknowledgments}
We are very grateful to Hui Li, Xilong Fan, and Bing Zhang for helpful discussions. This work is supported by the National Key R\&D Program of China (Nos. 2020YFC2201400, 2023YFC2205901), and the National Natural Science Foundation of China under grants 12473012 and 12533005. W.H.Lei. acknowledges support by the science research grants from the China Manned Space Project with NO.CMS-CSST-2021-B11.
\end{acknowledgments}

\bibliography{refereedoc}

\end{document}